\documentclass[twocolumn,showpacs,preprintnumbers,amsmath,amssymb]{revtex4}

\usepackage{graphicx}
\usepackage{dcolumn}
\usepackage{bm}
\usepackage{color}
\usepackage{hyperref}
\hypersetup{pdftitle={Destruction of superconductivity in disordered materials : a dimensional crossover},pdfauthor={O. Crauste, F. Couedo, L. Berge, C.A. Marrache-Kikuchi and L. Dumoulin}}

\begin{document}



\title{Destruction of superconductivity in disordered materials : a dimensional crossover}


\author{O. Crauste}
\author{F. Cou\"{e}do}
\author{L. Berg\'{e}}
\author{C.A. Marrache-Kikuchi}
\email{Claire.Marrache@csnsm.in2p3.fr}
\author{L. Dumoulin}
\affiliation{%
CSNSM, CNRS-UMR8609, Universit\'{e} Paris Sud, Bat. 108, 91405 Orsay Campus, France
}%


\date{\today}


\begin{abstract}
The disorder-induced Superconductor-to-Insulator Transition in amorphous Nb$_x$Si$_{1-x}$ two-dimensional thin films is studied for different niobium compositions $x$ through a variation of the sample thickness $d$. We show that the critical thickness $d_c$, separating a superconducting regime from an insulating one, increases strongly with diminishing $x$, thus attaining values of over 100 {\AA}. The corresponding phase diagram in the $(d,x)$ plane is inferred and related to the three-dimensional situation. The two-dimensional Superconductor-to-Insulator Transition well connects with the three-dimensional Superconductor-to-Metal Transition.
\end{abstract}








\maketitle

\emph{Introduction.-} In disordered systems, the electronic ground state is the result of a competition between Coulomb interactions, disorder, which eventually leads to localization of charge carriers, and, when relevant, superconductivity. In this conflict between antagonistic forces, dimensionality plays a special role and determines what ground states are allowed. Indeed, in three-dimensional systems, two distinct quantum phase transitions, the Metal-to-Insulator Transition (MIT) \cite{Belitz1994, Evers2008, Trivedi2012} and the Superconductor-to-Metal Transition, separate the three possible ground states ; by contrast, in two dimensions, the system can only exhibit a direct Superconductor-to-Insulator Transition (SIT), since metals are theoretically forbidden in the absence of strong electron-electron interactions \cite{Abrahams2001, Feigelman2007, Pokrovsky2010}. One important question is then to understand how the three ground states (superconducting, metallic and insulating) that are possible in bulk systems evolve when the thickness is reduced.
More specifically, is an initially three-dimensional superconducting system affected by a thickness reduction in an universal manner or does this effect depend on the initial strength of superconductivity?

The thickness-tuned SIT in thin alloy films provides an interesting way to address this question. Indeed, in bulk systems, the different ground states can be continuously explored through a change of stoichiometry
. For example, in our system of interest, amorphous Nb$_x$Si$_{1-x}$ (a-NbSi), bulk films ($d\gtrsim 1000$ \AA) are superconducting for $x\gtrsim 12.6\%$, metallic for $9\% \lesssim x\lesssim 12.6\%$, and insulating below $x\simeq9\%$ \cite{Dumoulin1993, Crauste2010, Bishop1985}. In the two-dimensional limit, the sample thickness $d$ is one of the parameters tuning the SIT : starting from a superconducting thin film, a reduction of $d$ progressively drives the system towards an insulating state, which it reaches below a critical thickness $d_c$ \cite{Haviland1989, Marrache2008}. For instance, pure niobium films ($x=100\%$) are superconducting until $d_c=7$ \AA \ \cite{Asamitsu1994, Nishida1991}. For a-NbSi alloys of lower niobium content, one can wonder whether $d_c$ depends on the composition $x$ and hence on the bulk superconducting temperature $T_{c0}$.

In the present letter, we focus on the thickness-induced SIT in two-dimensional a-NbSi films. We will present the variation of the critical thickness with the composition and consider the corresponding two-dimensional phase diagram in relation with the three-dimensional one.

\emph{Experimental.--}
a-NbSi films have been prepared at room temperature and under ultrahigh vacuum (typically a few 10$^{-8}$ mbar) by electron beam co-deposition of Nb and Si, at a rate of about 1 {\AA}.s$^{-1}$. The evaporation rates of each source were monitored \emph{in situ} by a dedicated set of piezo-electric quartz crystals in order to precisely monitor the composition and the thickness of the films during the deposition. These were also corroborated \emph{ex situ} by Rutherford Backscattering Spectroscopy (RBS) measurements \cite{Walls}. The films were deposited onto sapphire substrates coated with a 250 {\AA}-thick SiO underlayer designed to smooth the substrate surface. The samples were subsequently protected from oxidation by a 250 {\AA}-thick SiO overlayer. The samples studied here have Nb concentrations ranging from 13.5\% to 18\% and thicknesses varying from 20 to 500 {\AA}. Similar films have been measured to be continuous, amorphous and structurally non-granular at least down to a thickness of $d$ = 25 {\AA} \cite{Crauste2013}. The disorder scale in those films can therefore be estimated to be of the order of the inter-atomic distance.

Transport measurements were carried out down to 10 mK in a dilution refrigerator, using a resistance measurement bridge \cite{TRMC2} and standard AC lock-in detection techniques. The applied polarisation has been checked to be sufficiently low to be in the ohmic regime, or, when superconducting, below the critical current. All electrical leads were filtered from RF at room temperature.



\begin{figure}[h]
\begin{centering}
\includegraphics[width=0.9\columnwidth]{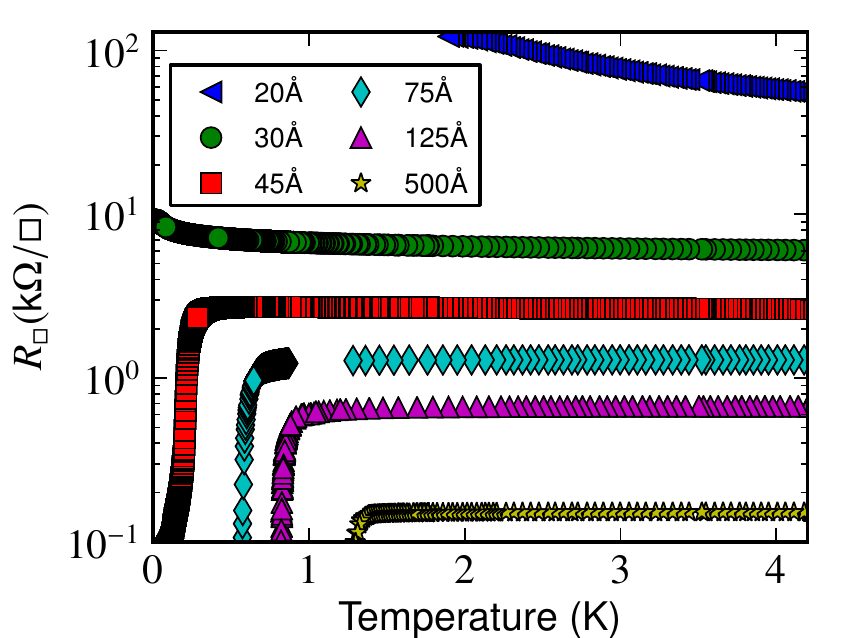}\\
  \caption{Sheet resistance as a function of temperature for a-Nb$_{18}$Si$_{82}$ samples of thicknesses ranging from 20 to 500 {\AA}. The SIT is tuned through a change in sample thickness and occurs at $d_c$ = 32 $\pm$ 1 \AA.}
  \label{fig:dc_encadrement_R(T)}
\end{centering}
\end{figure}

\emph{Determination of $d_c$.-} We have considered the evolution of the superconducting properties of two-dimensional a-NbSi thin films as the thickness is lowered, for different values of the composition $x$ (see figure \ref{fig:dc_encadrement_R(T)} for the different sheet resistances $R_\square(T)$ corresponding to $x$=18\%). The SIT occurs at a critical thickness $d_c$ characterized by a change of sign in the Temperature Coefficient of Resistance (TCR = $\frac{d\text{R}}{d\text{T}}$) at low temperature : a positive TCR corresponds to a superconducting ground state, whereas a negative TCR is taken to be characteristic of an insulator \cite{Fisher1990,  Schneider2012}. For each value of $x$, we have determined $d_c$ through two methods.

\begin{figure}[h]
\begin{centering}
\includegraphics[width=0.9\columnwidth]{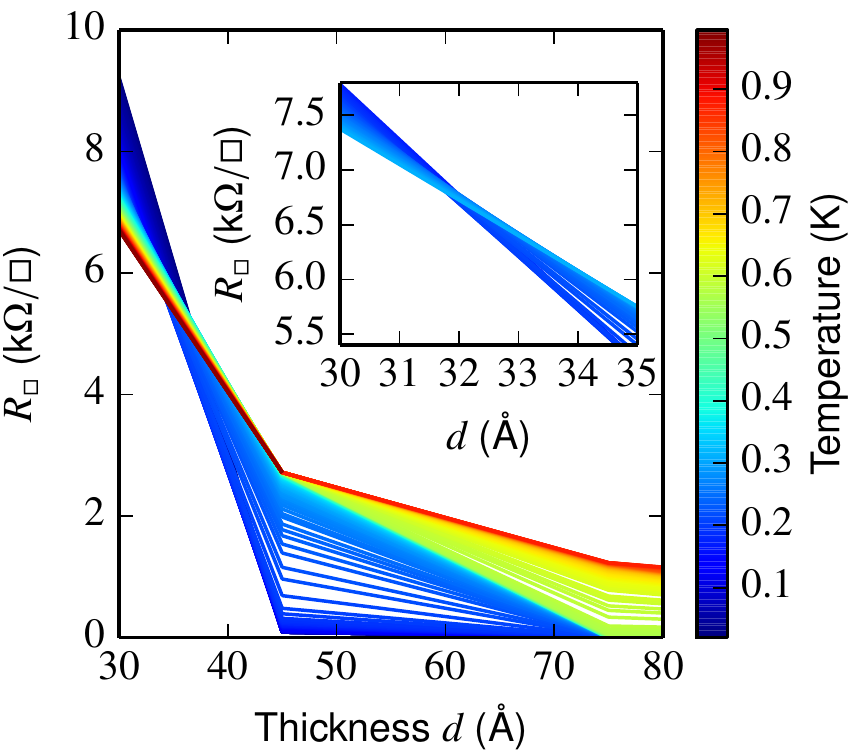}\\
  \caption{Sheet resistance as a function of the thickness for different temperatures ranging from 20 mK to 1 K for $x=18$\%. The crossing point provides $d_c$ and $R_{\square,c}$ at the transition. For $x=18\%$, $d_c=32\pm2$ \AA. Inset : same data, centered around $d=32$ \AA, for 170 mK $<$ $T$ $<$ 315 mK.}
  \label{fig:dc_scaling}
\end{centering}
\end{figure}

First, by plotting the evolution of the sheet resistance $R_\square$ with the thickness for different temperatures (see figure \ref{fig:dc_scaling}). The crossing point, $\left(d_c,R_{\square,c}\right)$, where $R_{\square,c}$ is the critical sheet resistance at the transition, signals the transition between a superconducting and an insulating phase \cite{Marrache2008, Hebard1990, Markovic1998}.

\begin{figure}[h]
\begin{centering}
\includegraphics[width=0.8\columnwidth]{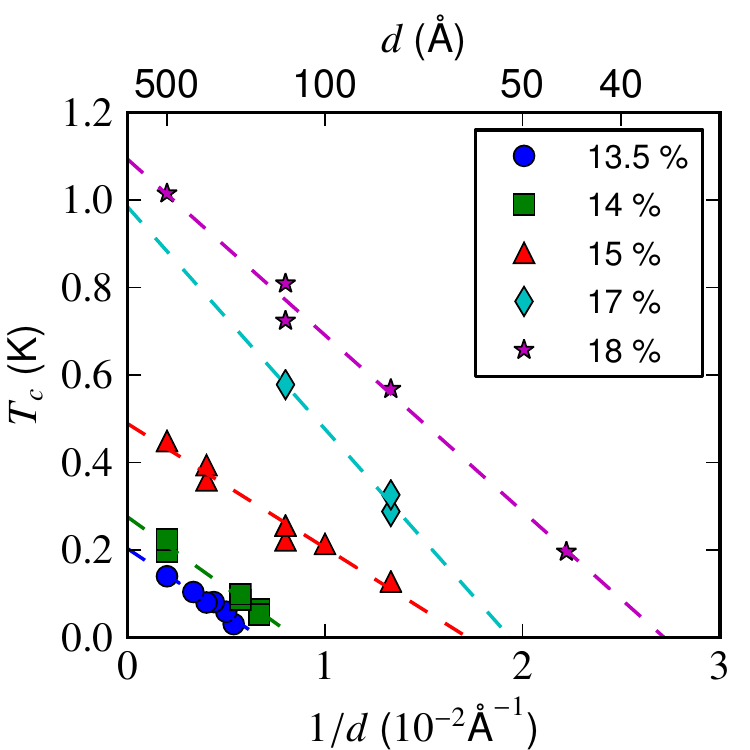}\\
  \caption{Evolution of the superconducting critical temperature $T_c$ with the inverse of the thickness $d$ for the different samples studied. The dashed lines are best linear fits to $T_c=f\left(1/d\right)$.}
  \label{fig:Tc_vs_d}
\end{centering}
\end{figure}

\begin{figure}[h]
\begin{centering}
\includegraphics[width=0.9\columnwidth]{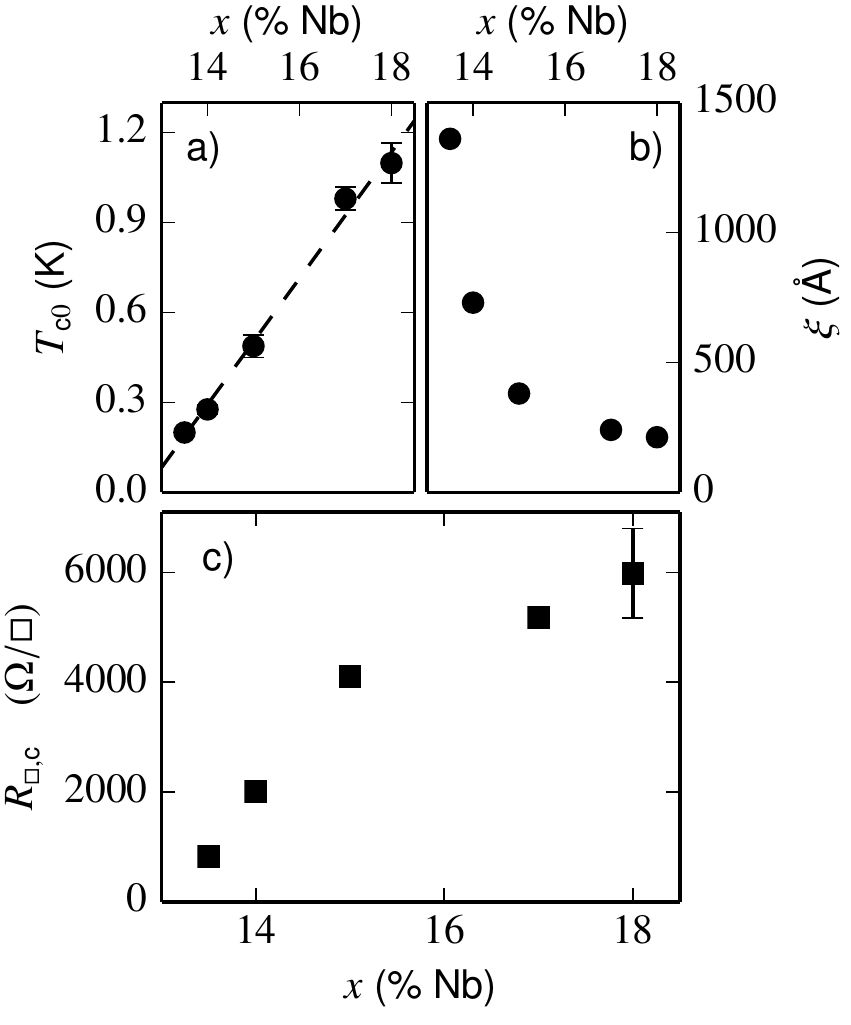}\\
  \caption{a. Bulk superconducting critical temperature ($T_{c0}$) for different compositions. The dotted line corresponds to the best linear fit : $T_{c0}=0.210(x-12.6)$. b. Evolution of the superconducting coherence length $\xi_{SC}$ with the composition $x$ (see text). c. Evolution with $x$ of the critical sheet resistance $R_{\square,c}$ estimated from the crossing point in the $R_\square(d)$ plot.}
  \label{fig:Tc0_xi_Rc}
\end{centering}
\end{figure}

The second estimation of $d_c$ derives from the evolution of the superconducting critical temperature $T_c$ with the thickness. $T_c$ is here taken to be the temperature below which $R_{\square}=0$ \cite{Note_Tc}. Indeed, as has been reported in other two-dimensional systems \cite{Stewart2009, Simonin1986}, at a given composition, we observe a relation between $T_c$ and $d$ : as the sample thickness is decreased, $T_c$ decreases linearly with $1/d$, as shown figure \ref{fig:Tc_vs_d}. The interpretation of this relation is beyond the scope of this paper. However, from the extrapolation to $1/d\rightarrow 0$, one can infer $T_{c0}$, the superconducting critical temperature of the bulk film corresponding to the same composition (figure \ref{fig:Tc0_xi_Rc}.a.). The values of $T_{c0}$ thus obtained are in very good agreement with what has previously been measured in bulk samples \cite{Juillard1996}. From the extrapolation to $T_c\simeq 0$, one can deduce the critical thickness at which superconductivity is suppressed in thin a-NbSi films.

\begin{figure}[h]
\begin{centering}
\includegraphics[width=0.9\columnwidth]{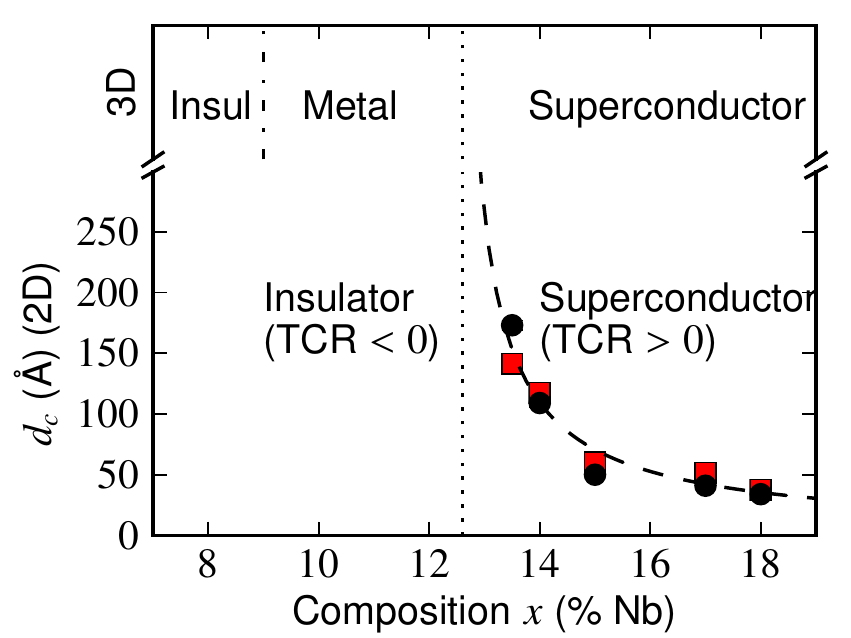}\\
  \caption{Phase diagram as a function of the composition $x$ and the thickness $d$. The black dots correspond to the estimation of $d_c$ as the value at which the $R_{\square}(d)$ curves cross at different temperatures. The red squares correspond to the value obtained by the extrapolation of $T_c(1/d)\rightarrow0$.  The dashed line corresponds to the best fit of the scaling law $d_c=f(x)$ (see text for details).}
  \label{fig:diag_phase}
\end{centering}
\end{figure}

Both evaluations of $d_c$ are given figure \ref{fig:diag_phase} and are in good agreement with one another. The obtained values of $d_c$ are large ($d_c>30$ \AA), which is particularly convenient to \emph{finely} study the disorder-tuned SIT at thicknesses where the continuity of the films is secured.


Even at these large values of $d_c$, the films can be considered two-dimensional from the point of view of superconductivity. Indeed, for a-NbSi films of a similar composition, the superconducting coherence length $\xi_{SC}$ has been measured to be larger than 100 {\AA} \cite{Pourret2006}. Moreover, a \emph{minimal} estimate of $\xi_{SC}$ can be derived from Gor'kov developments of the Ginzburg-Landau theory in the dirty limit : $\xi_{SC}=0.36\sqrt{\frac{3}{2}\frac{2\pi\hbar^3}{k_BT_{C0}mR_{\square,c} e^2}}$, where $m$ is the electron mass, $e$ its charge, and $R_{\square,c}$ the sheet resistance of the critical film \cite{Tinkham}. The dependance of this evaluation of $\xi_{SC}$ with $x$ is given figure \ref{fig:Tc0_xi_Rc}.b. We have $d\lesssim\xi_{SC}$, which is the commonly accepted criterion of 2D superconductivity \cite{Feigelman2012}, for all samples except those of thickness 500 {\AA}.

\emph{Critical resistance.-} In the literature, the value of the normal sheet resistance for the critical sample, $R_{\square,c} = \frac{h}{e^2\left(k_Fl\right)_c}$, has been deemed to be a direct measurement of the disorder $\left(k_Fl\right)_c$ at the SIT \cite{Fisher1990, Finkelstein1994}. In the bosonic scenario developed by M.P.A. Fisher, where Cooper pairs and vortices are related by a strict duality relation, the critical resistance has been predicted to be universal and of value the quantum resistance $R_Q=\frac{h}{4e^2}\simeq 6.5 \text{k}\Omega/\square$ \cite{Cha1991, Fisher1990}. Experimentally however, this universality is scarcely ever observed. Various explanations have been given for this discrepancy : $R_{\square,c}$ could be smaller than $R_Q$ due to weak spin-orbit interaction \cite{Bielejec2002} or to the contribution of fermionic excitations \cite{Yazdani1995}, but $R_{\square,c}$ could also be found to be larger than $R_Q$ \cite{Gantmakher1998, Liu1993} due to failure of the strict self-duality requirement of Fisher's theory.

The experimental evolution of $R_{\square,c}$ with the composition for a-NbSi thin films is given figure \ref{fig:Tc0_xi_Rc}.c. As can be seen, $R_{\square,c}$ is non universal \cite{Crauste2009} and varies by a factor of 7 for the considered composition range. The fact that $x$ can be tuned in a-NbSi films allows to span this large range of $R_{\square,c}$ in a single compound. It is interesting to note that, as $x$ increases above 12.6 \% - the value at which the superconductor-to-metal transition occurs in bulk samples - $R_{\square,c}$ evolves towards $R_Q$. This coincides with a SIT occurring at lower thicknesses, where the bosonic scenario, developed by M.P.A. Fisher, is more likely to be valid. Further investigation are needed to assess whether $R_{\square,c}$ reaches a maximum value of 6.5 k$\Omega$ or if it exceeds $R_Q$ in this system.


\emph{Phase diagram.-} Figure \ref{fig:diag_phase} summarizes the results under a phase diagram in the ($d$,$x$) plane, featuring the superconducting and insulating phases. Let us highlight some of the features of this diagram :

First, the critical thickness can be tuned by a variation in the film composition : $d_c$ seems to diverge as $x$ decreases. The corresponding evolution can be captured by a power law : $d_c=d_0\times\left(\frac{x-x_c}{x_c}\right)^\alpha$. Taking into account the uncertainty on the determination of $d_c$, the best fit (dashed line in figure \ref{fig:diag_phase}) gives $d_0=17\pm7$ {\AA}, $\alpha=-0.9\pm0.1$ and $x_c=12.4\pm0.6\%$. This equation  can be extrapolated to $x=100\%$ where it gives $d_c=5\pm2$ {\AA}, in agreement with what is found in the literature for pure Nb films \cite{Asamitsu1994, Nishida1991}.

Second, the critical thicknesses thus obtained can be as large as a few hundreds of angstr\"{o}ms. As far as we know, these are the largest critical thicknesses obtained for a two-dimensional SIT : $d_c$ is usually of a few monolayers for pure metal films \cite{Strongin1970, Haviland1989, Qin2009, Eom2006, Li2010} or a few tens of angstr\"{o}ms for other alloys \cite{Lee1990, Graybeal1984, Hirakawa2008}.

Third, the critical line separating the superconducting and insulating phases in two dimensions clearly extrapolates with the Superconductor-to-Metal boundary in the three-dimensional limit. Indeed, $x_c$, at which $d_c\rightarrow +\infty$, can be compared to the composition at which superconductivity ceases to exist in bulk films (linear fit of figure \ref{fig:Tc0_xi_Rc}.a : $T_{c0}=0$ for $x_{c0}=12.6\pm0.8\%$). Both values of $x_c$ coincide within error bars.

\emph{Discussion.-} The present work is therefore an original study of the variation of $d_c$ with a parameter driving the SIT in two dimensions, here the alloy composition $x$. Some hypothesis can be put forward to explain the remarkable result of the divergence of $d_c$ near a critical value of $x_{c0}$.

In the study of the thickness-induced destruction of superconductivity, there have been two different standpoints. The first, pioneered by Blatt and Thompson \cite{Blatt1963}, has emphasized the effect of surfaces on the confinement of the electronic wavefunctions. These so-called \emph{Quantum Size Effects} (QSE) are observable for very clean systems where the level spacing due to quantum confinement $\Delta E$ is larger than the scattering-induced broadening of the levels $\frac{\hbar}{\tau}$, where $\tau$ is the scattering time. QSE are believed to explain how $T_c$ oscillates with the thickness \cite{Eom2006, Ozer2007, Brun2009} in pure ultra-thin metallic films. In those cases, two atomic layers-thick films have been measured to present superconductivity, at least locally \cite{Qin2009}. Within this perspective, the critical thickness at which $T_c\rightarrow 0$ is expected to be $d_c=\frac{2a}{N(0)V}$ where $a$ is the Thomas-Fermi screening length \cite{Loptien2014}, and $N(0)V$ the electron-phonon coupling potential. It then is difficult to understand whether $d_c$ originates from a weakening of the electron-phonon coupling as the thickness is reduced or if it could be tuned through an engineering of the Fermi wave vector. The second standpoint is given by the SIT theories, which apply to disordered films where QSE should play a less prominent role. There again, it is not clear whether the suppression of superconductivity is due to enhanced superconducting fluctuations preventing a long range order to establish itself in \emph{any} film of reduced dimensionality - in which case $d_c$ would be independent of the material considered - , or whether $d_c$ depends on the density of states, and therefore on the initial strength of superconductivity.


a-NbSi films display large values of $d_c$ and are intrinsically disordered. This system therefore offers a chance to study the suppression of superconductivity without any QSE and study a possible correlation between $d_c$ and $T_{c0}$, which then is tunable via $x$. Compiling the experimental results obtained on pure metals does not permit to validate this hypothesis. Indeed, thickness reduction has most often been studied on superconducting materials with large $T_{c0}$, such as Pb \cite{Jaeger1989}, Nb \cite{Asamitsu1994}, a-Bi \cite{Haviland1989}. Metallic films with $T_{c0}$ lower than 1 K (W, Ir, Ti, Al \cite{Strongin1970}) exhibit, before the destruction of superconductivity, an increase of $T_c$ when the film thickness is reduced. This entangles the impact of disorder with an increase of surface effects on the phonon spectrum \cite{Belitz1987}, making experiments more complicated to interpret. The tunability of $T_{c0}$ through $x$, without additional surface effects, is specific to alloys such as a-NbSi, and has enabled us to carry out a systematic study of the SIT critical thickness with $T_{c0}$ \emph{within the same material}. In the present study, $x$ only varies from 13.5\% to 18\% in a structurally disordered compound, so that films of different compositions are very similar, material-wise. A correlation between $T_{c0}$ and $d_c$, if confirmed, would be an original result, calling for an in-depth study.

Another hypothesis is to relate the divergence of $d_c$ with the proximity of the MIT in the corresponding three-dimensional material. Indeed, bulk a-NbSi presents anomalies in the density-of-state, related to the correlation pseudo-gap that develops near the MIT \cite{Bishop1985}. It would be therefore interesting to link the divergence of $d_c$ to the opening of this pseudo-gap. This assumption will be tested in future tunneling experiments.

Regarding the continuity in the phase diagram between the bulk metal and the two-dimensional insulating state - characterized, as is usual in the literature, by a negative TCR - the crucial issues then are : first to understand what microscopic differences exist between those two states. For instance, the nature of the two-dimensional insulator is still a debated question \cite{Trivedi2012}. Second : how does the three-dimensional metal transit to an insulating ground state at lower dimension and at what lengthscale. Further investigations on this point are under way.

\emph{Conclusion.-} We have studied the thickness-tuned SIT in amorphous Nb$_x$Si$_{1-x}$ thin films and established the phase diagram when the composition of the alloy is varied. The critical thickness below which the system is insulating is correlated with the bulk superconducting critical temperature, and diverges in the vicinity of the critical composition at which the Superconductor-to-Metal Transition occurs in bulk films. The tunability of a-NbSi, and notably the possibility of modifying its bulk superconducting critical temperature, makes it a model system to study the SIT.





\begin{acknowledgments}
 We gratefully thank Vincent Humbert for his careful reading of the manuscript. This work has been partially supported by the Agence Nationale de la Recherche (grants No. ANR-06-BLAN-0326 and ANR-2010-BLANC-0403-01), and by the Triangle de la Physique (grant No. 2009-019T-TSI2D).
\end{acknowledgments}





\begin{thebibliography}{47}
\bibitem[{Not()}]{Note_Tc}
\bibinfo{note}{Evaluations of $T_c$ as the temperature at which the TCR is
  maximum, at which $R_{\square}=R_n/2$ - with $R_n$ the normal sheet
  resistance -, or below which $R_{\square}=0$, give the same results with a
  maximum discrepency of a few percents.}
\end{thebibliography}
\end{document}